\documentclass[a4paper,12pt]{article}

\usepackage{amsmath}
\usepackage{amssymb}
\usepackage{amsfonts}
\usepackage{amsthm}
\usepackage{color}
\usepackage{cancel} 
\usepackage{ulem}
\usepackage{cite}

\newcommand{\be}{\begin{equation}}
\newcommand{\ee}{\end{equation}}
\newcommand{\bea}{\begin{eqnarray}}
\newcommand{\eea}{\end{eqnarray}}
\newcommand{\bref}[1]{(\ref{#1})}
\newcommand{\nn}{\nonumber}

\newcommand{\mapright}[1]{\smash{[mathop{\hbox to 1cm{\rightarrowfill}}
\limits^{#1}}}

\newcommand{\del}{\partial}

\allowdisplaybreaks[3]
\def\dd{\hbox{\,\large$\rhd$}}		

 \makeatletter
 \@addtoreset{equation}{section}
 \makeatother

\date{empty}
\begin{document}
\begin{titlepage}
\null
\begin{flushright}
December, 2022 \\
KEK-TH-2482
\end{flushright}
\vskip 2cm
\begin{center}
{\Large \bf 
Gauged Double Field Theory, Current Algebras
\\
\vspace{0.3cm}
and Heterotic Sigma Models
}
\vskip 1.5cm
\normalsize
\renewcommand\thefootnote{\alph{footnote}}

{\large
Machiko Hatsuda${}^{*}$\footnote{mhatsuda(at)juntendo.ac.jp}, 
Haruka Mori${}^{\dagger}$\footnote{h.mori(at)sci.kitasato-u.ac.jp},
Shin Sasaki${}^{\dagger}$\footnote{shin-s(at)kitasato-u.ac.jp}
and 
Masaya Yata${}^{\ddagger}$\footnote{m-yata(at)juntendo.ac.jp}
}
\vskip 0.5cm
  {\it
  ${}^{*}$
  Department of Radiological Technology, Faculty of Health Science,
  Juntendo University \\
  \vspace{-0.1cm}
  Yushima, Bunkyo-ku, Tokyo 113-0033, Japan \\
  ${}^{*}$KEK Theory Center, High Energy Accelerator Research Organization \\
  \vspace{-0.1cm}
  Tsukuba, Ibaraki 305-0801, Japan \\
  ${}^{\dagger}$
  Department of Physics, Kitasato University, 
  Sagamihara 252-0373, Japan \\
  ${}^{\ddagger}$
  Physics Division, Faculty of Medicine, Juntendo University, Inzai, Chiba 270-1695, Japan
  }
\vskip 1.0cm
\begin{abstract}
We study the $O(D,D+n)$ generalized metric and the
 gauge symmetries in the gauged double field theory (DFT)
in view of current algebras and sigma models.
We show that the $O(D,D+n)$ generalized metric in 
the gauged DFT is consistent with
 the heterotic sigma models at the leading order in the $\alpha'$-corrections.
We then study the non-Abelian gauge symmetries and current algebras of heterotic string theories.
We show that the algebras exhibit the correct diffeomorphism, the $B$-field
 gauge transformations of the background fields together with the
 non-Abelian gauge transformations possibly with the appropriate local Lorentz
 transformations.
\end{abstract}
\end{center}
\end{titlepage}

\newpage
\setcounter{footnote}{0}
\renewcommand\thefootnote{\arabic{footnote}}
\pagenumbering{arabic}
\tableofcontents

\section{Introduction} \label{sect:introduction}

Dualities are key ingredients to understand the overall picture of string theories.
Among other things, dualities in heterotic string theories \cite{Hull:1994ys} have
been attracted much attention due to their sophisticated gauge structures.
Heterotic string theories consist of the $D=26$ left moving bosonic string together
with the $D=10$ right moving superstring \cite{Gross:1985fr,Gross:1984dd}.
The consistency of the theories requires the gauge groups to be $SO(32)$
or $E_8 \times E_8$.
These structures are inherited by the low energy effective theories of
heterotic strings -- the heterotic supergravities.
Due to the famous anomaly cancellation mechanism \cite{Green:1984sg}, it is necessary to
include the $\alpha'$- and also the higher derivative corrections in the
heterotic supergravities \cite{Bergshoeff:1988nn, Bergshoeff:1989de}.
The first order corrections of $\alpha'$ are given in the form of the
gauge kinetic term together with the Riemann curvature squares
and the Chern-Simons completion of the modified field strength of the
$B$-field.
It is noteworthy that the Riemann curvature square terms in the heterotic
supergravities contain the fourth orders of spacetime derivatives.
Therefore they are $\alpha'$- and also the derivative corrections.
Determining $\alpha'$- and derivative corrections to supergravities is
not straightforward in general.

Type II supergravities compactified on the torus $T^D$ have T-duality group $O(D,D)$.
It is shown that $O(D,D)$ structures are preserved to all orders in $\alpha'$ \cite{Sen:1991zi}.
The higher derivative corrections to
supergravities are discussed \cite{Eloy:2020dko} and it is shown that 
$\alpha'$- and derivative corrections are highly restricted by the
$O(D,D)$ structures.

Double field theory (DFT) \cite{Hull:2009mi, Hull:2009zb} developed based on the doubled formalism
\cite{Tseytlin:1990nb, Siegel:1993xq, Siegel:1993th, Siegel:1993bj}, is
a reformulation of supergravities that keeps manifest $O(D,D)$
structures.
DFT is utilized to study various aspects of string theories.
Recent developments are focused on the $\alpha'$-corrections in the
doubled formalism \cite{Hohm:2013jaa, Hohm:2014xsa}.
A further development, the so-called gauged double field theory
\cite{Hohm:2011ex, Grana:2012rr} including $\alpha'$-corrections
\cite{Bedoya:2014pma, Lescano:2021guc} enable us to write down the $O(D,D+n)$ covariant
formulation of heterotic supergravities.
A universal formulation of $\alpha'$-corrections in the $O(D,D)$
covariant formalism has been developed \cite{Marques:2015vua}.
Among other things, the $O(D,D+n)$ structure is utilized to embed the
derivative corrections to heterotic supergravities.
It is shown that the Lorentz spin connection in the 
generalized metric in DFT precisely reproduces the Riemann curvature
square term under the section condition \cite{Bedoya:2014pma}.
This exhibits further evidence for the availability of duality to
determine the derivative corrections in supergravities.

The purpose of this paper is to make connections between the heterotic
sigma models and the gauged DFT including the non-Abelian gauge
symmetries, $\alpha'$- and derivative
corrections. We will show that the $O(D,D+n)$ generalized metric in
\cite{Bedoya:2014pma} is derived from a heterotic sigma model
in arbitrary backgrounds supplemented by an extended internal space.
Rather than the fermionic formulation of heterotic sigma models utilized
in \cite{Sen:1985qt, Sen:1985eb}, we consider the equivalent bosonic formulation of
the theory \cite{Narain:1986am}. 
We will also discuss the current algebras for the heterotic sigma model
and show that the algebras are consistent with the non-Abelian gauge transformations
of the background fields.

The organization of this paper is as follows.
In the next section, we introduce the generalized metric for heterotic
supergravities in the language of double field theory.
The doubled formalism requires the internal space spanned by
the $n$-dimensional gauge space.
In section \ref{sect:sigma_models}, we reproduce heterotic sigma models
from the enlarged generalized metric in the gauged DFT.
We also discuss the current algebras of the heterotic theory.
We find that the algebra exhibits non-Abelian gauge transformations of the background gauge fields.
We finally comment on a possible extension to include derivative corrections.
We point out that the curvature square term in DFT and heterotic
supergravities are implemented in the sigma model with the extra 
unphysical fields.
Section \ref{sect:conclusion} is devoted to conclusion and discussions.


\section{Heterotic supergravities and double field theory} \label{sect:generalized_metric}
In this section, we briefly introduce heterotic supergravities and
exhibit the structure of the $\alpha'$- and derivative corrections.
We also introduce the $O(D,D+n)$ generalized metric and the (gauged) DFT that
reproduces heterotic supergravities.

\subsection{Heterotic supergravities}
The action for the bosonic sector of heterotic supergravities in
$\mathcal{O} (\alpha')$ is given by
\begin{align}
S =& \ \frac{1}{2 \kappa^2_{10}} \int \! d^{10} x \sqrt{-g} e^{-2\phi}
\Big[
R (\omega) - \frac{1}{12} \hat{H}_{\mu \nu \rho} \hat{H}^{\mu \nu \rho} +
 4 \del_{\mu} \phi \del^{\mu} \phi
\notag \\
& \qquad \qquad \qquad \qquad \qquad
- \frac{\alpha'}{4}
\Big(
\mathrm{Tr} F_{\mu \nu} F^{\mu \nu} + R_{\mu \nu mn} (\omega_-) R^{\mu
 \nu mn} (\omega_-)
\Big)
\Big] .
\label{eq:heterotic_sugra_action}
\end{align}
Note that the ratio of the gravitational constant $\kappa_{10}$ and the gauge
coupling constant $g_{10}$ is $\frac{\kappa^2_{10}}{2 g^2_{10}} = \alpha'$.
Here $\phi$ is the dilation and $F_{\mu \nu}$ is the field strength of
the $SO(32)$ or $E_8 \times E_8$ gauge field $A_{\mu}$.
The metric $g_{\mu \nu}$ is defined by the vielbein $e_{\mu} {}^m$
as $g_{\mu \nu} = \eta_{mn} e_{\mu} {}^m e_{\nu} {}^n$ where $\eta_{mn}$
is the flat metric in the local Lorentz frame.
The Riemann tensor $R^{mn} {}_{\mu \nu}$ and the 
Ricci scalar $R (\omega)$ are defined by the spin connection
$\omega_{\mu} {}^{mn}$;
\begin{align}
R^{mn} {}_{\mu \nu} (\omega) =& \ \del_{\mu} \omega_{\nu} {}^{mn}
- \del_{\nu} \omega_{\mu} {}^{mn} 
+ \omega_{\mu} {}^{mq} \eta_{pq} \omega_{\nu} {}^{pn} 
- \omega_{\nu} {}^{mq} \eta_{pq} \omega_{\mu} {}^{pn},
\notag \\
R (\omega) =& \ e^{\mu} {}_m e^{\nu} {}_n R^{mn} {}_{\mu \nu} (\omega).
\label{eq:curvature}
\end{align}
The modified field strength $\hat{H}_{\mu \nu \rho}$ of the $B$-field
$B_{\mu \nu}$ is given by 
\begin{align}
\hat{H}_{\mu \nu \rho} = H_{\mu \nu \rho} + \alpha'
 (\Omega^{\mathrm{YM}}_{\mu \nu \rho} - \Omega^{L-}_{\mu \nu \rho}),
\end{align}
where $H = d B$ and $\Omega^{\mathrm{YM}}$, $\Omega^{L-}$ are the
Chern-Simons terms associated with the Yang-Mills and the Lorentz
connections \cite{Bergshoeff:1988nn};
\begin{align}
\Omega^{\mathrm{YM}}_{\mu \nu \rho} =& \  3! \mathrm{Tr}
\Big(
A_{[\mu} \del_{\nu} A_{\rho]} + \frac{2}{3} A_{[\mu} A_{\nu} A_{\rho]}
\Big),
\notag \\
\Omega^{L-}_{\mu \nu \rho} =& \ 3! 
\Big(
\eta_{np} \eta_{mq} \omega_{-[\mu} {}^{mn} \del_{\nu} \omega_{- \rho]}
 {}^{pq}
+ \frac{2}{3} \eta_{ms} \eta_{np} \eta_{qr} \omega_{- [\mu} {}^{mn}
 \omega_{- \nu} {}^{pq} \omega_{- \rho]} {}^{rs}
\Big).
\label{eq:Chern-Simons_terms}
\end{align}
Here and in the following, the totally anti-symmetrization symbol $[\cdots]$ includes the
weight factor $\frac{1}{n!}$.
Note that the modified spin connection $\omega_{\pm \mu} {}^{mn}$ is
defined by 
\begin{align}
\omega_{\pm \mu} {}^{mn} = \omega_{\mu} {}^{mn} \pm \hat{H}_{\mu} {}^{mn}.
\end{align}
Here $\hat{H}_{\mu} {}^{mn} = e^{\nu m} e^{\rho n} \hat{H}_{\mu \nu
\rho}$.
Since the spin connection $\omega_{\mu} {}^{mn}$ is given by 
\begin{align}
\omega_{\mu} {}^{mn} =& \ 
\frac{1}{2}
\Big[
e^{m \nu} (\del_{\mu} e_{\nu} {}^n - \del_{\nu} e_{\mu} {}^n)
-
e^{n \nu} (\del_{\mu} e_{\nu} {}^m - \del_{\nu} e_{\mu} {}^m)
-
e^{m \rho} e^{n \sigma} (\del_{\rho} e_{\sigma p} - \del_{\sigma}
 e_{\rho p}) e_{\mu} {}^p
\Big],
\end{align}
the last term in the action \eqref{eq:heterotic_sugra_action} is in the
fourth order of the derivative expansion.
The action is correct up to the two derivatives of the gauge field and $\mathcal{O} (\alpha')$, and invariant under
the following gauge transformations;
\begin{align}
\delta B_{\mu \nu} =& \ 2 \del_{[\mu} \tilde{\xi}'_{\nu]} - \alpha' 
\Big\{
\kappa_{\alpha \beta} \xi^{\alpha} F^{\beta}_{\mu \nu} 
- 
\kappa_{\Lambda \Sigma} \xi^{\Lambda} R^{\Sigma}_{\mu \nu}
\Big\},
\notag \\
\delta A_{\mu \beta} =& \ \del_{\mu} \xi^{\alpha} \kappa_{\alpha \beta} + f^{\gamma}
 {}_{\alpha \beta} A_{\mu \gamma} \xi^{\alpha},
\notag \\
\delta \omega_{- \mu \Lambda} =
& \ 
\del_{\mu} \xi^{\Sigma} \kappa_{\Sigma \Lambda} + f^{\Gamma}
 {}_{\Lambda \Sigma}\ \omega_{-\mu \Gamma} \xi^{\Lambda},
\label{eq:sugra_gauge}
\end{align}
where $\kappa_{\alpha \beta}, \kappa_{\Lambda \Sigma}$ and 
$f^{\alpha} {}_{\beta \gamma}, f^{\Gamma} {}_{\Lambda \Sigma}$
 are the Cartan-Killing forms and the structure constants of the heterotic gauge groups
and the local Lorentz group.
Here 
$\tilde{\xi}_{\mu}', \xi^{\alpha}, \xi^{\Sigma}$ are 
the gauge parameters for the $B$-field gauge symmetry, 
the $SO(32)$ or $E_8 \times E_8$ gauge transformations and the local
Lorentz transformations.
The gauge indices $\alpha, \beta, \ldots$, the local Lorentz indices
$\Sigma, \Lambda, \ldots = [mn]$ and the component of 
the field strength $F^{\alpha}_{\mu \nu}$ for $A_{\mu}^{~\alpha}$ have
been introduced.
The gauge and the local Lorentz indices are raised and lowered by the Cartan-Killing forms
$\kappa_{\alpha \beta}, \kappa_{\Lambda \Sigma}$ and their inverses.
Note that $R_{\mu \nu} {}^{\Lambda}$ is the Riemann tensor defined in
\eqref{eq:curvature}.

\subsection{Generalized metric and gauged double field theory}
The action \eqref{eq:heterotic_sugra_action} at
$\mathcal{O} (\alpha')$ is obtained via the $O(D,D+n)$ invariant double
field theory \cite{Hohm:2011ex, Grana:2012rr, Bedoya:2014pma}.
Since the Riemann curvature square term contains spacetime derivatives
of the fourth order, we first focus on
the second derivative action except this term.
We also ignore the Lorentz Chern-Simons term $\Omega^{L-}$ for a moment.
The doubled coordinate $X^M \, (M=1, \ldots, 2D+n)$ is decomposed into the spacetime $X^{\mu}$, the 
winding $\tilde{X}_{\mu}$ and the internal coordinates $Y_{\alpha}$,
{\it i.e.} $X^M = (\tilde{X}_{\mu}, Y_{\alpha}, X^{\mu})$, where $\mu,
\nu = 1, \ldots, D$, $\alpha = 1, \ldots, n$. 
The $Y_{\alpha}$ directions are responsible for a gauge group of
dimension $n$.
The $O(D,D+n)$ invariant metric is defined by
\begin{align}
\eta_{MN}=
\left[
    \begin{array}{ccc}
      0  &  0   &  \delta^{\mu}_{~\nu} \\  
      0  &  \kappa^{\alpha\beta}  &  0  \\
      \delta_{\mu}^{~\nu}  & 0 &  0 
    \end{array}
  \right],
\qquad
\eta^{MN}=
\left[
    \begin{array}{ccc}
      0  &  0   &  \delta_{\mu}^{~\nu} \\  
      0  &  \kappa_{\alpha\beta}  &  0  \\
      \delta^{\mu}_{~\nu}  & 0 &  0 
    \end{array}
  \right],
\label{eq:invariant_metric}
\end{align}
where the $n \times n$ matrices $\kappa_{\alpha \beta}, \kappa^{\alpha
\beta}$ are the Cartan-Killing form and its inverse for the $SO(32)$ or $E_8 \times E_8$ gauge groups.
The $O(D,D+n)$ (enlarged) generalized metric is parametrized by 
\begin{align}
{\cal H}^{MN} &= 
\left[
    \begin{array}{ccc}
      g_{\mu\nu} +\alpha' A_{\mu}^{\alpha} A_{\nu\alpha} + c_{\rho\mu} g^{\rho\sigma} c_{\sigma\nu}  &  \sqrt{\alpha'}A_{\mu\alpha} +\sqrt{\alpha'}A_{\rho\alpha} g^{\rho\sigma} c_{\sigma\mu}   &  -c_{\rho\mu}g^{\rho\nu} \\  
      \sqrt{\alpha'}A_{\nu\beta} + \sqrt{\alpha'} A_{\rho\beta} g^{\rho\sigma} c_{\sigma\nu}  &  \kappa_{\alpha\beta} + \alpha' A_{\rho\alpha} g^{\rho\sigma} A_{\sigma\beta} &  -\sqrt{\alpha'} A_{\rho\beta} g^{\rho\nu}  \\
      -c_{\sigma\nu} g^{\sigma\mu}  & -\sqrt{\alpha'} A_{\sigma\alpha} g^{\sigma\mu} &  g^{\mu\nu}
    \end{array}
  \right],
\label{eq:heterotic_generalized_metric}
\end{align}
where $g_{\mu \nu}$, $B_{\mu \nu}$ are symmetric and 
anti-symmetric tensors while $A_{\mu}^{\alpha}$ is a vector.
The $c_{\mu \nu}$ is defined by 
$c_{\mu \nu} = B_{\mu \nu} + \frac{\alpha'}{2} A_{\mu}^{\alpha} A_{\nu
\alpha}$.
All the quantities depend on $X^M$.
The generalized metric $\mathcal{H}^{MN}$ and its inverse $\mathcal{H}_{MN}$
satisfy the $O(D,D+n)$ relation $\mathcal{H}^{MN} = \eta^{MP} \eta^{NQ}
\mathcal{H}_{PQ}$.
Note that the $O(D,D+n)$ indices are raised and lowered by the
$O(D,D+n)$ invariant metrices \eqref{eq:invariant_metric}.
The action of the so-called gauged double field theory is given by \cite{Hohm:2011ex}
\begin{align}
S_{\mathrm{gDFT}} =& \ \int \! d^{2D + n} X \, e^{-2d} 
\Bigg[
\,
\frac{1}{8} \mathcal{H}^{MN} \del_M \mathcal{H}^{KL} \del_N
 \mathcal{H}_{KL}
- \frac{1}{2} \mathcal{H}^{MN} \del_N \mathcal{H}^{KL} \del_L
 \mathcal{H}_{MK}
\notag \\
& \qquad \qquad \qquad \qquad 
- 2 \del_M d \del_N \mathcal{H}^{MN} + 4 \mathcal{H}^{MN} \del_M d
 \del_N d
\notag \\
& \qquad \qquad \qquad \qquad 
- \frac{1}{2} f^M {}_{NK} \mathcal{H}^{NP} \mathcal{H}^{KQ} \del_P
 \mathcal{H}_{QM}
- \frac{1}{12} f^M {}_{KP} f^N {}_{LQ} \mathcal{H}_{MN} \mathcal{H}^{KL}
 \mathcal{H}^{PQ}
\notag \\
& \qquad \qquad \qquad \qquad 
- \frac{1}{4} f^M {}_{NK} f^N {}_{ML} \mathcal{H}^{KL}
- \frac{1}{6} f^{MNK} f_{MNK}
\,
\Bigg],
\label{eq:gauged_DFT_action}
\end{align}
where $e^{-2d} = \sqrt{-g} e^{-2\phi}$ is the $O(D,D+n)$ invariant generalized dilaton and
$\phi$ is a scalar function of $X^M$.
The constant $f^M {}_{NP}$ satisfies
\begin{align}
f^{(M} {}_{PK} \eta^{N)K} = 0,
\qquad
f^M {}_{N[K} f^N {}_{LP]} = 0.
\end{align}
The second one is the Jacobi identity.
The action \eqref{eq:gauged_DFT_action} is invariant under the
 following gauge transformations;
\begin{align}
\delta_{\xi} \mathcal{H}^{MN} =& \ 
\xi^P \del_P \mathcal{H}^{MN} 
+ 
\left(
\del^M \xi_P - \del_P \xi^M
\right) \mathcal{H}^{PN}
+
\left(
\del^N \xi_P - 
\partial_P 
\xi^N
\right) \mathcal{H}^{MP}
- 2 \xi^P f^{(M} {}_{PK} \mathcal{H}^{N)K},
\notag \\
\delta_{\xi} d =& \ \xi^M \del_M d - \frac{1}{2} \del_M \xi^M,
\label{eq:DFT_gauge_symmetry}
\end{align}
provided the conditions
\begin{align}
&
\eta^{MN} \del_M \del_N * = \eta^{MN} \del_M * \del_N * = 0,
\notag \\
& 
f^M {}_{NK} \del_M * = 0,
\label{eq:section_constraints}
\end{align}
are satisfied. Here $*$ is the generalized metric, the generalized dilaton and the
gauge parameters.
These conditions are also derived by the closure of the 
algebra.
For example, the commutator of the transformation
\eqref{eq:DFT_gauge_symmetry} results in
\begin{align}
[\delta_{\xi_1}, \delta_{\xi_2}] V^M 
&= \delta_{[\xi_1, \xi_2]_f} V^M + T^M (\xi^1, \xi^2, V) \notag\\
&\quad - \eta^{MR} \xi_{1}^{P} f^{Q}{}_{RP} (\partial_{Q} \xi_{2N}) V^{N} 
		+ \eta^{MR} \xi_{2}^{P} f^{Q}{}_{RP} (\partial_{Q} \xi_{1N}) V^{N} \notag\\
&\quad  + f^{L}{}_{NK} (\partial_{L} \xi_{1}^{M}) \xi_{2}^{N}  V^{K}
		- f^{L}{}_{NK}  (\partial_{L} \xi_{2}^{M}) \xi_{1}^{N} V^{K} 
		- \ \xi_{2}^{N}\xi_{1}^{K} f^{P}{}_{NK} \partial_{P} V^{M}\notag\\
&\quad + \frac{1}{2} \eta^{MR}\xi_{1}^{L} f^{N}{}_{RK} (\partial_{N} \xi_{2L}) V^{K}
		- \frac{1}{2} \eta^{MR}\xi_{2}^{L} f^{N}{}_{RK} (\partial_{N} \xi_{1L}) V^{K}
,
\label{Cbra1}
\end{align}
where $V^M$ is set to be $E_A{}^M$ 
and the twisted C-bracket 
is defined by \cite{Hohm:2011ex, Mori:2020yih}
\begin{align}
[\xi_1, \xi_2]_f^{M} &= \xi_1^K \del_K \xi_2^M - \xi_2^K \del_K \xi_1^M - \frac{1}{2} \eta^{MN}
 \eta_{KL} (\xi_1^K \del_N \xi_2^L - \xi_2^K \del_N \xi_1^L)
+ \xi_2^N \xi_1^K f^M {}_{NK}.
\label{Cbra2}
\end{align}
The term $T^M$ is given by
\begin{align}
T^M (\xi^1, \xi^2, V) &= \frac{1}{2} \eta_{KL}
\big(
\xi_1^K \del^P \xi_2^L - \xi_2^K \del^P \xi_1^L
\big) \del_P V^M
-
\big(
\del^P \xi_1^M \del_P \xi_2^K - \del^P \xi_2^M \del_P \xi_1^K
\big) V_K.
\end{align}
It is apparent that the gauge algebra closes, on the twisted
C-bracket, by the conditions \eqref{eq:section_constraints}.
We call the conditions \eqref{eq:section_constraints} the section
constraints.
%

For $SO(32)$ or $E_8 \times E_8$ heterotic theories in ten dimensions,
we have $D=10$ and $n = 496$ and the covariant structure constant $f^M
{}_{NK}$ is chosen such as 
\begin{align}
f^M {}_{NK} = 
\left\{
\begin{array}{ll}
f^{\alpha} {}_{\beta \gamma} & \text{if } (M,N,K) = (\alpha, \beta,
 \gamma) 
 \\
0 & \text{else}
\end{array}
\right.
\end{align}
where $f^{\alpha} {}_{\beta \gamma}$ is the structure constant of the 
$G = SO(32)$ or $G = E_8 \times E_8$ gauge group. 
This choice breaks the global $O(D,D+n)$ invariance down to that of $O(D,D) \times
G$. 
Note that the section constraints \eqref{eq:section_constraints} are trivially
solved by $\tilde{\del}^{\mu} * = \del_{\alpha} * = 0$. 
This implies that all the quantities depend only on $x^{\mu}$.
In this case, the gauged DFT action \eqref{eq:gauged_DFT_action} reduces to the
one for the heterotic supergravity \eqref{eq:heterotic_sugra_action}
(except the Riemann tensor square term).
The component fields $g_{\mu \nu}, B_{\mu \nu}, \phi, A_{\mu}^{~\alpha}$ are
then identified with the spacetime metric, the NSNS $B$-field, the dilaton
and the $SO(32)$ or $E_8 \times E_8$ gauge field.
With this setup, the DFT gauge symmetry
\eqref{eq:DFT_gauge_symmetry} reduces to the gauge symmetry
\eqref{eq:sugra_gauge}.
Specifically, the $O(D,D+n)$ gauge parameter $\xi^M$ is decomposed into
those for the diffeomorphism $\xi^{\mu}$, the $B$-field gauge symmetry
$\tilde{\xi}_{\mu}$ and the non-Abelian gauge symmetry $\xi^{\alpha}$.
Note that the $O(D,D+n)$ covariant expression
\eqref{eq:DFT_gauge_symmetry} ensures the mixing of the $B$-field and
the non-Abelian gauge symmetries in \eqref{eq:sugra_gauge}.

The $O(D,D+n)$ generalized metric $\mathcal{H}^{MN}$ is expressed by the
generalized vielbein $E_{A}{}^{M}$,
\begin{align}
{\cal H}^{MN} &=E_A{}^{M} \hat{\eta}^{AB} E_{B}{}^{N},
\end{align}
where $\hat{\eta}^{AB}$ is an $H$-invariant metric.
We use the following parametrizations;
\begin{align}
E_{A}{}^{M} &=   
  \left[
    \begin{array}{ccc}
      e_{\mu}^{~m}  &  0   &  0 \\  
      \sqrt{\alpha'} A_{\mu}^{a}  &  e_{\beta}^{a}  &  0  \\
      -c_{m\mu}  &  -\sqrt{\alpha'} A_{m\beta}  &  e_m^{~\mu} 
    \end{array}
  \right], 
~~~~~
\hat{\eta}^{AB}=
\left[
    \begin{array}{ccc}
      \eta_{mn}  &  0   &  0 \\  
      0  &  \eta_{ab}  &  0  \\
      0  & 0 &  \eta^{mn} 
    \end{array}
  \right],
\end{align}
where $a,b, \ldots$ are the flat tangent space indices in the gauge direction.
Here $e_{\mu}^{~m}$ satisfies $g_{\mu \nu} = e_{\mu}^{~m} e_{\nu}^{~n}
\eta_{mn}$ and is identified with the spacetime vielbein when the
section constraints are imposed.
The flat space metric\footnote{
We consider the Euclidean signature rather than the Minkowski one.
The latter is straightforwardly obtained by switching the first $+1$ to
$-1$ in $\eta_{mn}$.
}
 is given by $\eta_{mn} = \mathrm{diag} (+1,+1, \ldots,+1)$ 
where $m,n = 1, \ldots, D$ are the flat space indices.
The flat metric $\eta_{ab}$ is numerically equivalent to $\kappa_{\alpha \beta}$ and 
the index $a = 1, \ldots, n$ is assumed to run over the gauge degrees of freedom.
The generalized vielbein $E_{A}{}^{M}$ satisfies
\begin{align}
E_A{}^ME_B{}^N\eta_{MN}=\eta_{AB},
\end{align}
where 
\begin{align}
\eta_{AB}=
\left[
    \begin{array}{ccc}
      0  &  0   &  \delta^{m}_{~n} \\  
      0  &  \eta^{ab}  &  0  \\
      \delta_{m}^{~n}  & 0 &  0 
    \end{array}
  \right].
\end{align}
The Cartan-Killing form $\kappa_{\alpha \beta}$ is given by the
gauge space vielbein $e_{\alpha}^{a}$ and the flat metric $\eta_{ab}$ as 
$\kappa_{\alpha \beta}= \eta_{ab}  e^{a}_{\alpha} e^{b}_{\beta}$. 

In the frame formalism, the gauge transformation is given by
\begin{align}
\delta_{\xi} E_{A}{}^{M} = \xi^P \del_P E_{A}{}^{M} + (\del^M \xi_P - \del_P \xi^M)
 E_{A}{}^{P} - f_{PQ} {}^M \xi^P E_{A}{}^{Q}.
\label{eq:gauge_transformation_vielbein}
\end{align}
With the transformation of the generalized dilaton $d$ in
\eqref{eq:DFT_gauge_symmetry} and the section constraints, we have the
transformation rule for the component fields \cite{Bedoya:2014pma};
\begin{align}
\delta \phi =& \ \xi^{\rho} \del_{\rho} \phi,
\notag \\
\delta e_{\mu}^{~m} =& \ \xi^{\rho} \del_{\rho} e_{\mu}^{~m} +
 \del_{\mu} \xi^{\rho} e_{\rho}^{~m},
\notag \\
\delta e_{\alpha}^{~a} =& \ 
\xi^{\rho} \del_{\rho} e_{\alpha}^{~a} - f_{\alpha \gamma}
 {}^{\beta} \xi^{\gamma} e_{\beta}^{~a},
\notag \\
\delta A_{\mu}^{~\beta} =& \ 
\xi^{\rho} \del_{\rho} A_{\mu}^{~\beta} + \del_{\mu} \xi^{\rho}
 A_{\rho}^{~\beta}
+ \del_{\mu} \xi^{\beta} - f_{\alpha \gamma} {}^{\beta} \xi^{\alpha}
 A_{\mu}^{~\gamma},
\notag \\
\delta B_{\mu \nu} =& \ 
\xi^{\rho} \del_{\rho} B_{\mu \nu}
- 2 \del_{[\mu} \xi^{\rho} B_{\nu]\rho}
+ 2 \del_{[\mu} \tilde{\xi}_{\nu]}'
+ \alpha' \del_{[\mu} A_{\nu]}^{~\alpha} \kappa_{\alpha \beta}
 \xi^{\beta},
\label{eq:gauge_transformation_components}
\end{align}
where we have rescaled the gauge parameters 
$\xi^M \to (\xi_{\mu}, \sqrt{\alpha'} \xi_{\alpha}, \xi^{\mu})$
and redefined as 
\begin{align}
\tilde{\xi}'_{\mu} = \xi_{\mu} - \frac{\alpha'}{2} 
A_{\mu}^{~\alpha} \kappa_{\alpha \beta} \xi^{\beta}.
\label{redef:xi}
\end{align}
Then we find that (\ref{eq:gauge_transformation_components}) is consistent with the gauge transformations (\ref{eq:sugra_gauge}) in heterotic supergravities.

\section{$O(D,D+n)$ generalized metric and heterotic sigma model} \label{sect:sigma_models}
In this section, we discuss relations among the $O(D,D+n)$ generalized
metric $\mathcal{H}^{MN}$, heterotic sigma models,
the gauge symmetries and current algebras.
In the reformulation of heterotic supergravity in terms of the gauged
DFT, the internal gauge directions $Y_{\alpha}$ have been assumed to be
496 dimensions \cite{Hohm:2011ex, Grana:2012rr}.
This is necessary in order to implement the non-Abelian gauge groups
$SO(32)$ or $E_8 \times E_8$ in DFT.
However, in the original formulation of heterotic string theories, the
dimension of the internal torus should be 16 since the central charges
of the left moving bosonic and the right moving superstring must cancel.
This dimension coincides with that of the Cartan subgroup $U(1)^{16}$ of $SO(32)$ or
$E_8 \times E_8$.
In the following, we start with the gauge directions $Y_{\alpha}$ of
dimension 16 ($\alpha = 1, \ldots, 16$) for the $U(1)^{16}$ gauge
background and then examine a non-Abelian generalization of the background.

\subsection{Generalized metric and sigma model}
Following the usual procedure of sigma models for the bosonic string in arbitrary
backgrounds, we assume that the Hamiltonian $H$ in the two-dimensional heterotic sigma
model is given by
\begin{align}
H = { 1 \over \sqrt{-h} h^{00} } H_{\tau} - { h^{01} \over h^{00}
 }H_{\sigma}, \label{Ham}
\end{align}
where $h^{ij} \, (i,j=0,1)$ is the two-dimensional worldsheet metric and each part is
expressed by
\begin{align}
H_{\sigma} = { 1 \over 2} \dd_M \eta^{MN} \dd_N,
\qquad
H_{\tau} = { 1 \over 2} \dd_M {\cal H}^{MN} \dd_N.
\label{1.9}
\end{align}
They are Virasoro operators associated with the two-dimensional worldsheet directions
$\sigma^i = (\tau, \sigma)$.
Here $\dd_M$ stands for the basis of the ``stringy derivatives'' that acts on
background fields (see the discussions below).
This is decomposed as
\begin{align}
\dd_M =
\left[
    \begin{array}{c}
      \dd^{\mu}   \\  
      \dd^{\alpha}    \\
      \dd_{\mu}   
    \end{array}
  \right]
=
\left[
    \begin{array}{c}
      \partial X^{\mu}   \\  
      \dd^{\alpha}    \\
      P_{\mu}   
    \end{array}
\right]. \label{currents}
\end{align}
The components $(\dd^{\mu},\dd_{\mu})$ are given by $(\partial X^{\mu}, P_{\mu})$
as in the case of the ordinary bosonic string sigma models.
Here $\del X^{\mu} = \del_{\sigma} X^{\mu}$ is the derivative of the
spacetime coordinate $X^{\mu}$ and $P_{\mu}$ is the momentum of the string.
For the remaining component $\dd^{\alpha}$, 
which is a derivative in the gauge space, 
we define a linear combination of the internal gauge space coordinates
$\del Y_{\alpha}$ and the momentum $\Pi_{\alpha}$ conjugate to $Y_{\alpha}$;
\begin{align}
\dd^{\alpha} = {1\over\sqrt{2}} (\partial Y^{\alpha} + \kappa^{\alpha\beta} \Pi_{\beta}).
\label{currentY} 
\end{align}
The Hamiltonian in \bref{Ham} with \bref{1.9}, \bref{currents}, \bref{currentY} 
in the $O(D,D+n)$ background \bref{eq:heterotic_generalized_metric}  is given with $\lambda_0=(\sqrt{-h} h^{00})^{-1}$ and
$\lambda_1=h^{01}/h^{00}$  as
\bea
H&=&{\lambda_0\over2} g^{\mu\nu} P_{\mu} P_{\nu} 
+P_\mu\partial X^\nu\{\lambda_0g^{\mu\rho}(B_{\rho\nu}-\frac{\alpha'}{2}A_\rho{}^\alpha A_\nu{}_\alpha)-\lambda_1\delta_\nu^\mu\}
\nn\\
&&+{\lambda_0\over2} \partial X^{\mu} \partial X^{\nu}
 ( g_{\mu\nu} + B_{\rho\mu}
B_{\sigma\nu} g^{\rho\sigma} + \alpha' A_{\mu}^{\alpha} A_{\nu\alpha}
+\alpha'B_{\mu\rho}g^{\rho\lambda}A_{\lambda}{}^{\alpha} A_{\nu\alpha}
 ) \nn\\
&&+{1\over4}(\Pi_\alpha+\partial Y_\alpha)(\Pi_\beta+\partial Y_\beta)
\{(\lambda_0-\lambda_1) \kappa^{\alpha\beta}+\alpha'\lambda_0 g^{\mu\nu}
A_\mu{}^\alpha A_\nu{}^\beta\}
\nn\\
&&
-\sqrt{\alpha'\over2}
\lambda_0 P_\mu (\Pi_\alpha+\partial Y_\alpha)g^{\mu\nu} A_\nu{}^\alpha
+
\sqrt{\alpha'\over2}
\lambda_0 (\Pi_\alpha+\partial Y_\alpha) \partial X^\mu A_\nu{}^\alpha
(g^{\nu\rho} B_{\rho\mu}-\delta_\mu^\nu)~~~.\label{Haminbg}
\eea
The Lagrangian is given by
\bea
{\cal L}&=&P_\mu \dot{X}^\mu+\Pi_\alpha \dot{Y}^\alpha -H\nn\\
&=&\frac{1}{2}\sqrt{-h}h^{ij}\partial_i X^\mu \partial_j X^\nu g_{\mu\nu}
-\frac{1}{2}\epsilon^{ij}\partial_i X^\mu \partial_j X^\nu B_{\mu\nu}
\nn
\\
&&+\frac{1}{\lambda_0-\lambda_1}(\dot{Y}^\alpha \dot{Y}_\alpha -
\sqrt{2\alpha'} \dot{Y}^\alpha \dot{X}^\mu A_\mu{}_\alpha
+\frac{\alpha'}{2}\dot{X}^\mu\dot{X}^\nu A_\mu{}^\alpha A_\nu{}_\alpha )\nn\\&&
-(\dot{Y}^\alpha \partial{Y}_\alpha -
\sqrt{2\alpha'} \dot{Y}^\alpha \partial{X}^\mu A_\mu{}_\alpha
+\frac{\alpha'}{2}\dot{X}^\mu\partial{X}^\nu A_\mu{}^\alpha
A_\nu{}_\alpha ),~~~
\label{Lag1}
\eea
with $i=0,1$. 
The string action is given as $I= - \int {\cal L}$.
The first and the second terms in \eqref{Lag1}
are covariant with respect to the worldsheet Lorentz symmetry. 
 The worldsheet non-covariance of the $Y^\alpha$ direction
is caused from  the chiral property of $\dd_\alpha$.
It is important to bear in mind that the internal fields $Y^{\alpha}$ in 
heterotic string theories in flat space obey the chirality condition
$\dot{Y}^{\alpha} - \del Y^{\alpha} = 0$.
In the presence of the background, the chirality condition of
$Y^{\alpha}$ is replaced as
\bea
&&\partial_- Y^\alpha=\dot{Y}^\alpha-\partial Y^\alpha=0~~\nn\\
&&\Rightarrow ~~
J_-{}^\alpha=\partial_- Y^\alpha-\sqrt{\frac{\alpha'}{{2}}}\partial_- X^\mu A_\mu{}^\alpha
=j_-{}^\alpha+ \tilde{j}_-{}^\alpha=0~~, \nn \\
&&~~~~~~J_\pm{}^\alpha~=~J_0{}^\alpha\pm J_1{}^\alpha~~,~~
j_i{}^\alpha=\partial_i Y^\alpha~~,~~
\tilde{j}_i{}^\alpha=
-\sqrt{\frac{\alpha'}{{2}}}\partial_i X^\mu A_\mu{}^\alpha~~~.
\label{chiralitycondition}
\eea
We add the
square of the chirality condition $\lambda (J_-{}^\alpha)^2$
to
 the Lagrangian \bref{Lag1} in such a way that the kinetic term allows to take the conformal gauge \cite{Siegel:1983es, Hatsuda:2018tcx, Hatsuda:2019xiz}.
The positivity of $\kappa_{\alpha\beta}$ leads the condition
 $ (J_-{}^\alpha)^2=0$ to $J_-{}^\alpha=0$.
Here $\lambda$ is the Lagrange multiplier.
Then the terms including $Y^{\alpha}$ and $A_{\mu}^{\alpha}$ in the
Lagrangian \eqref{Lag1} become
\bea
{\cal L}_{\rm gauge}&=&\frac{1}{\lambda_0-\lambda_1}(\dot{Y}^\alpha \dot{Y}_\alpha -
\sqrt{2\alpha'} \dot{Y}^\alpha \dot{X}^\mu A_\mu{}_\alpha
+\frac{\alpha'}{2}\dot{X}^\mu\dot{X}^\nu A_\mu{}^\alpha A_\nu{}_\alpha )\nn\\
&&-(\dot{Y}^\alpha \partial{Y}_\alpha -
\sqrt{2\alpha'} \dot{Y}^\alpha \partial{X}^\mu A_\mu{}_\alpha
+\frac{\alpha'}{2}\dot{X}^\mu\partial{X}^\nu A_\mu{}^\alpha A_\nu{}_\alpha )
+\lambda J_-{}^\alpha J_-{}_\alpha\nn\\
&=&\frac{1}{\lambda_0-\lambda_1} J_0{}^\alpha J_0{}_\alpha
-J_0{}^\alpha J_1{}_\alpha+\lambda J_-{}^\alpha J_-{}_\alpha
+\sqrt{\frac{\alpha'}{2}}\epsilon^{ij}\partial_i Y^\alpha \partial_j X^\mu A_{\mu\alpha}
~~~\nn\\
&=&\frac{1}{4}
\left\{\left(\frac{1}{\lambda_0-\lambda_1}-1\right) J_+{}^\alpha J_+{}_\alpha
+\frac{2}{\lambda_0-\lambda_1}J_+{}^\alpha J_-{}_\alpha
+\left(\frac{1}{\lambda_0-\lambda_1}+1+4\lambda\right)J_-{}^\alpha J_-{}_\alpha
 \right\}\nn\\
&&+\epsilon^{ij}j_i{}^\alpha  \tilde{j}_j{}_\alpha
~~~.\label{Lagpm}
\eea
In the following gauge 
\bea
&&\lambda_0=1~~,~~\lambda_1=0~~,~~\lambda=-\frac{1}{2}~~~,\label{gaugechoice}
\eea
the Lagrangian \bref{Lagpm} becomes the conformal form in the worldsheet as
\bea
{\cal L}_{\rm gauge}~=~\frac{1}{2}J_+{}^\alpha J_-{}_\alpha
+\epsilon^{ij}j_i{}^\alpha  \tilde{j}_j{}_\alpha
~~~.\label{Laggauge}
\eea
Then the Lagrangian \bref{Lag1} in the gauge \bref{gaugechoice} becomes
\bea
{\cal L}&=&\frac{1}{2}\partial_+ X^\mu \partial_- X^\nu g_{\mu\nu}
-\frac{1}{2}\epsilon^{ij}\partial_i X^\mu \partial_j X^\nu B_{\mu\nu}+\frac{1}{2}J_+{}^\alpha J_-{}_\alpha
+\epsilon^{ij}j_i{}^\alpha  \tilde{j}_j{}_\alpha
~~~\nn\\
&=&\frac{1}{2}\partial_+ X^\mu \partial_- X^\nu g_{\mu\nu}
-\frac{1}{2}\epsilon^{ij}\partial_i X^\mu \partial_j X^\nu B_{\mu\nu}
+\frac{1}{2}\partial_+Y^\alpha \partial_-{Y}_\alpha+\sqrt{\frac{\alpha'}{2}}\epsilon^{ij}\partial_i Y^\alpha \partial_j X^\mu A_{\mu\alpha}\nn\\
&&
-\sqrt{\frac{\alpha'}{8}}
(\partial_+Y^\alpha \partial_-{X}^\mu A_\mu{}_\alpha
+\partial_-Y^\alpha \partial_+{X}^\mu A_\mu{}_\alpha)
+\frac{\alpha'}{4}
 \partial_+{X}^\mu A_\mu^\alpha  \partial_-{X}^\nu A_\nu{}_\alpha~~~
.\label{conformalLagrangian}
\eea

The first line of the Lagrangian is nothing but
the one for the bosonic sector of heterotic string theories
\cite{Narain:1986am}.
The first and second terms in the second line are cross terms of the bilinear of the currents $J_+{}^\alpha J_-{}_\alpha$.
The last term includes the next order of $\alpha'$ in the Lagrangian. 
The conjugate momentum 
is derived from the Lagrangian 
\bref{conformalLagrangian} as
\bea
\Pi_\alpha=\displaystyle\frac{\partial {\cal L}}{\partial \dot{Y}^\alpha}
=\dot{Y}_\alpha-\sqrt{\frac{\alpha'}{{2}}}\partial_- X^\mu A_\mu{}_\alpha=\partial Y_\alpha~~\Rightarrow~~
\Pi_\alpha-\partial Y_\alpha=0~~~,
\eea
showing that the right-moving mode is 0 with the chirality condition
\bref{chiralitycondition}.
The Lagrangian 
\bref{conformalLagrangian} as
is therefore consistent in this order $\mathcal{O} (\sqrt{\alpha'})$.

\subsection{Unphysical coordinates and gauge symmetries}
As we have mentioned, the background gauge fields
$A_{\mu}^{~\alpha}$ 
correspond to the Cartan subsector of the heterotic gauge
groups and therefore $\alpha = 1, \ldots, 16$ as same as \cite{Narain:1986am}.
However the gauge group of the $O(D,D+n)$ DFT is promoted to
the full non-Abelian heterotic gauge groups $A_{\mu}^{~\alpha} \, (\alpha = 1, \ldots,
496)$ by the gauging. 
In order to fill this gap, let us recall how to incorporate the non-Abelian gauge symmetries in heterotic string theories.
A heterotic string consists of the supersymmetric right-moving part and
the left-moving part describing the gauge symmetry.
The ten-dimensional coordinate $X^\mu(\tau,\sigma)$ includes both the left and the right-moving modes.
In order to cancel the Virasoro anomaly, the left-moving part must also include
the 16 internal bosonic coordinates 
$Y^{\hat{\alpha}} \, (\hat{\alpha} = 1,\ldots, 16)$ for a torus $T^{16}$.
The coordinates $Y^{\hat{\alpha}}$ are $U(1)$ parameters of the torus $T^{16}$.

The non-Abelian symmetry is realized by the bosonized description \cite{Friedan:1985ge}.
For the $SO(32)$ group
the dimension of the Cartan subalgebra $H^{\hat{\alpha}}$ is 16,
while the dimension of the remaining generators $E_{K}$ is 480.
A set of free 16 left-moving bosons $Y^{\hat{\alpha}}$ gives all $so(32)$ currents.
The  $H^{\hat{\alpha}}$-current is represented as $\partial_+ Y^{\hat{\alpha}}$.
The  $E_{K}$-current is represented as $e^{iK\cdot Y}$ 
where $K_{\underline{\alpha}}\cdot Y=\pm Y^{\hat{\alpha}}\pm Y^{\hat{\beta}}, \,
(\hat{\alpha} \neq \hat{\beta})$ for a root vector of $SO(32)$, $K_{\underline{\alpha}}$. 
The number of $K_{\underline{\alpha}}$ is $_{16}C_{2}\times 2^2=480$.
This is generalized to the other gauge symmetry by choosing the
momentum vector $K_{\underline{\alpha}}$ belonging to a root lattice of the
corresponding gauge group.
In this way, the 496-dimensional generators of $so(32)$ are realized in
terms of just the 16 physical bosons $Y^{\hat{\alpha}}$.
However these currents are non-linear in coordinates $Y^{\hat{\alpha}}$.

We now turn to the discussion of the $O(D,D+n)$ covariant formulation.
Instead of introducing the 16 internal coordinates $Y^{\hat{\alpha}}$,
we introduce 496 coordinates representing 496 currents in the covariant
expression.
We will then eliminate the unphysical degrees of freedom by constraints.
We utilize the matrix valued coordinates studied in AdS string 
\cite{Roiban:2000yy, Hatsuda:2001xf}.
The $SO(32)$ coordinate is a 32$\times$32 matrix  $Z\in SO(32)$
satisfying $Z^TZ=1$. 
The left-invariant 1-form $j=Z^{-1}dZ=j_\alpha (i T^\alpha )$ 
is an element of the $so(32)$ algebra.
Here $T^{\alpha}$ are the generators of the $so(32)$ algebra.
It satisfies the Maurer-Cartan equation $dj=-j\wedge j $ which is rewritten as 
\bea
dj^{\alpha}&=&\frac{1}{2}f_{\beta\gamma}{}^\alpha j^{\beta}\wedge j^{\gamma}~
\eea
with the $so$(32) structure constant $f_{\alpha\beta}{}^{\gamma}$ for  $\left[T_\alpha,T_\beta\right]=if_{\alpha\beta}{}^{\gamma}T_\gamma$.
$Z(\sigma)$ is the left-moving coordinate, but we consider both left and right-moving modes and 
the chirality condition is imposed in the end.
The worldsheet current is $j_\pm{}^\alpha=\frac{1}{32i}{\rm tr}[(Z^{-1}\partial_\pm Z) ~T^\alpha]\in so(32)$ for the 32$\times$32 matrix representation of $T^\alpha$.
Physical currents are only 16 left-moving currents $j_+{}^{\hat{\alpha}}$,
while 480 left-moving currents $j_+{}^{\underline{\alpha}}$ and all right-moving currents $j_-{}^\alpha$
are unphysical.

With this fact in mind, we again focus on the Lagrangian \bref{conformalLagrangian} especially on \bref{Laggauge}.
Now we extend the gauge directions to 496 dimensions;
$\alpha=(\hat{\alpha},~\underline{\alpha})$, $({\hat{\alpha}}= 1, \ldots, 16,
~{\underline{\alpha}}= 17, \ldots, 496)$.
The Lagrangian contains the 480 extra ``unphysical'' scalar fields $Y^{\underline{\alpha}}$.
%
The gauge field coupling generalizes $j_\pm{}^\alpha\to J_\pm{}^\alpha$.
 Physical currents are only 16 left-moving currents $J_+{}^{\hat{\alpha}}$,
 while 480 left-moving currents $J_+{}^{\underline{\alpha}}$ and all right-moving currents $J_-{}^\alpha$ are unphysical. 
In addition to the constraint term  $\lambda (J_{-}^{\alpha})^2$
we further add $\lambda' (J_{+}^{\underline{\alpha}})^2$
 to \bref{Lag1}, then the Lagrangian becomes 
\bea
{\cal L}_{\rm gauge}
&=&\frac{1}{4}
\left\{
\left(\frac{1}{\lambda_0-\lambda_1}-1\right) J_+{}^{\hat{\alpha}} J_+{}_{\hat{\alpha}}
+\left(\frac{1}{\lambda_0-\lambda_1}-1+\lambda'\right) J_+{}^{\underline{\alpha}} J_+{}_{\underline{\alpha}}
+\frac{2}{\lambda_0-\lambda_1}J_+{}^\alpha J_-{}_\alpha\right.\nn\\
&&\left.
+\left(\frac{1}{\lambda_0-\lambda_1}+1+4\lambda\right)J_-{}^\alpha J_-{}_\alpha
\right\}
+\epsilon^{ij}j_i{}^\alpha  \tilde{j}_j{}_\alpha\nn\\
&=&\frac{1}{2}J_+{}^\alpha J_-{}_\alpha
+\epsilon^{ij}j_i{}^\alpha \tilde{j}_{j\alpha}
+\lambda'J_{+}^{\underline{\alpha}}J_{+\underline{\alpha}}~~~.
\label{NAacionv2} 
\eea
where the gauge choice in \bref{gaugechoice} is used in the last line.
We propose a Lagrangian for the compactified 16-dimensional left-moving
space as the non-Abelian generalization of the gauge sector in \eqref{conformalLagrangian}.

\subsection{Current algebras}
We next discuss the gauge transformations of the backgrounds.
Current algebras of strings and branes have been utilized to determine the
gauge transformations of background fields \cite{Hatsuda:2020buq}.
The non-Abelian current algebra is obtained as follows.
The canonical conjugate of the $SO(32)$ coordinate $Z_{ij}$, $i,j=1,\ldots,32$,  ~$Z^TZ=1$ is introduced as
\bea
\left[P^{ij}(\sigma),Z_{kl}(\sigma')\right]&=&2i\delta_{[k}^i\delta_{l]}^j\delta(\sigma-\sigma')~,
\eea
where $P^{ij}$ is the conjugate momentum of $Z_{ij}$.
The covariant derivative is given as
\cite{Hatsuda:2001xf}
\bea
D_{ij}&=&(PZ)_{ij}-(PZ)_{ji}~\in~so(32)~.
\eea
Cartan subalgebras are $H^{\hat{\alpha}}=D_{[(2\hat{\alpha}-1)\ 
2\hat{\alpha}]}$ with $\hat{\alpha}=1,\cdots,16$, while  remaining generators are 
$E_{K}$.
The covariant derivative $D_{ij}$ and the $\sigma$-component of the left-invariant current $j_{1;ij}$ satisfy 
the following algebra
\bea
\left[D_{ij}(\sigma),D_{kl}(\sigma')\right]&=&4i \delta_{[l|[i}D_{j]|k]}\delta(\sigma-\sigma')\nn~,\\
\left[D_{ij}(\sigma),j_{1;kl}(\sigma')\right]&=&4i \delta_{[l|[i}j_{1;j]|k]}\delta(\sigma-\sigma')
+2i\delta_{l[i}\delta_{j]k} \partial_\sigma\delta(\sigma-\sigma')~.
\nn
\eea
The left-moving covariant derivative $\dd_{ij}=(D+j_{1})_{ij}$ satisfies
the following algebra 
by using $D-j_{1}=0$ which is imposed as the second class constraint on the right hand side 
\bea
\left[\dd_{ij}(\sigma),\dd_{kl}(\sigma')\right]&=&
6i\delta_{[l|[i}\dd_{j]|k]}\delta(\sigma-\sigma')
+4i\delta_{l[i}\delta_{j]k} \partial_\sigma\delta(\sigma-\sigma')\label{SO32}~.
\eea
The non-Abelian gauge transformations are implemented by
generalizing the current algebra \bref{SO32} as   
\begin{align}
	[\dd_\alpha(\sigma), \dd_\beta(\sigma')] = if_{\alpha\beta}{}^{\gamma}\dd_\gamma ~\delta(\sigma-\sigma') + i\kappa_{\alpha\beta} \partial_{\sigma} \delta(\sigma-\sigma')
\end{align}
where 
$\dd_{ij} = (iT^{\alpha})_{ij} \dd_{\alpha}$ and
$\kappa_{\alpha\beta}={\rm tr}(T_\alpha T_\beta)$ is the
non-degenerate Cartan-Killing form with the scale redefinition.

The $O(D,D+n)$ covariant stringy currents including the non-Abelian gauge generators are then given 
 as $\dd_M=(\dd^\mu,~\dd^\alpha,~\dd_\mu)$.
They satisfy the following current algebra;
\begin{align}
[\dd_M(\sigma), \dd_N(\sigma')] = if_{MN}{}^{K}\dd_K
 ~\delta(\sigma-\sigma') + i\eta_{MN} \partial_{\sigma}
 \delta(\sigma-\sigma')
\label{affineLA} 
\end{align}
with $f_{MN}{}^L=f_{\alpha\beta}{}^\gamma$ and $\eta_{\alpha\beta}=\kappa_{\alpha\beta}$.
The metric $\eta_{MN}$ must be non-degenerate and the group structure must be totally anti-symmetric 
\bea
\eta_{MN}=\left(
\begin{array}{ccc}&&\delta^\mu{}_\nu\\&\kappa^{\alpha\beta}&\\\delta_\mu{}^\nu&&
\end{array}
\right)~,~
f_{MN}{}^K\eta_{KL}\equiv f_{MNL}=\frac{1}{3!}f_{[MNL]}
\eea
 in order to satisfy the Jacobi identity of \bref{affineLA}.
The $O(D,D+n)$ enlarged space coordinates are 
$X^M = (\tilde{X}_{\mu}, Y_{\alpha}, X^{\mu})$
, 
$\mu=1,\cdots,10$ and $\alpha=1,\cdots,496$.
Derivative operations of a function on the space $\Phi(X^M)$ are given by 
\bea
\partial_M \Phi(X)&=&i\displaystyle\int  d\sigma' \left[\dd_M(\sigma'), \Phi(X)\right]~,\nn\\
\partial_\sigma \Phi(X)&=&i\displaystyle\int  d\sigma' \left[H_\sigma(\sigma'), \Phi(X)\right]=\dd_N\eta^{NM}(\partial_M \Phi)~.
\eea
The section conditions of fields $\Phi(X)$ and $\Psi(X)$ are given by
\bea
\partial_M\eta^{MN}\partial_N\Phi(X)&=&[2\partial^\mu\partial_\mu+(\partial^\alpha)^2]\Phi(X)=0,\nn\\
\partial_M\Phi(X)\eta^{MN}\partial_N\Psi(X)&=&
\partial^\mu\Phi\partial_\mu\Psi+
\partial_\mu\Phi\partial^\mu\Psi+
\partial^\alpha\Phi
\partial_\alpha\Psi
=0~.
\eea

Now the generalized vielbein
$E_A{}^M(X)$ includes the gauge field $A_\mu{}^\alpha(X)$ as well as the 
vielbein $e_m{}^{\mu}$ and 
the $B$-field $B_{\mu\nu}$. 
The vielbein
satisfies the orthogonal condition 
with respect to the $O(D,D+n)$ invariant metric
\bea
E_A{}^ME_B{}^N\eta_{MN}=\eta_{AB}.
\eea 
The transformation rules for the vielbein field $E_A{}^{M}$ can be written by the commutator;
\begin{align}
\delta E_A{}^M\dd_M &=  i \int d\sigma'[ \xi^M\dd_M(\sigma') , E_A{}^{N}\dd_N(\sigma)] \nonumber\\
                  &= \Bigl(   E_A{}^{M} (\partial_M \xi^N -
 \partial^N \xi_M)  +  \xi^M \partial_M E_A{}^{N}
 +f_{MK}{}^{N} E_A{}^{M}\xi^K\Bigr)\dd_N~. 
\label{eq:current_gauge_transformations_vielbein}
\end{align}
By imposing the section conditions and writing down each component in
\eqref{eq:current_gauge_transformations_vielbein}, we obtain 
\begin{align}
\delta e_{\beta}^{a} &=  \xi^{\mu} \partial_{\mu}
 e^{a}_{\beta} 
+ f^{\delta}_{~\gamma\beta} e^{a}_{\delta} \xi^{\gamma}, \nonumber\\
\delta e_m^{\mu} &= - e_m^{\nu} \partial_{\nu} \xi^{\mu} + \xi^{\mu}\partial_{\mu}e_m^{\nu}, \nonumber\\
\delta A_{\mu\beta} &=  \partial_{\mu} \xi^{\nu} A_{\nu\beta} 
+ \partial_{\mu} \xi^{\alpha} \kappa_{\alpha\beta} + \xi^{\nu}\partial_{\nu} A_{\mu\beta} +f^{\gamma}_{~\alpha\beta}A_{\mu\gamma}\xi^{\alpha}, \nonumber\\
\delta c_{\nu\mu} &=  c_{\rho\mu} \partial_{\nu} \xi^{\rho}  + c_{\nu\rho} \partial_{\mu} \xi^{\rho} + \xi^{\rho} \partial_{\rho} c_{\nu\mu} + (\partial_{\nu} \xi_{\mu}
 -\partial_{\mu} \xi_{\nu}) + 
\alpha' A_{\nu\beta} \partial_{\mu} \xi^{\beta},
\label{eq:current_gauge_transformations}
\end{align}
where we have rescaled the gauge parameter $\xi^{\alpha} \to
\sqrt{\alpha'} \xi^{\alpha}$
and assumed 
that all the quantities depend only on $x^{\mu}$.
This is always guaranteed by the section conditions.
From the last two lines in \eqref{eq:current_gauge_transformations}
, we can obtain the transformation rule for the $B$-field,
\begin{align}
\delta B_{\nu\mu} &=   \xi^{\rho} \partial_{\rho} B_{\nu\mu}   + ( B_{\rho\mu} \partial_{\nu} \xi^{\rho} - B_{\rho\nu} \partial_{\mu} \xi^{\rho})
                          + (\partial_{\nu} \xi_{\mu} - \partial_{\mu} \xi_{\nu})
 +
\frac{\alpha'}{2}
( A_{\nu\beta} \partial_{\mu} \xi^{\beta} - A_{\mu\beta} \partial_{\nu}
 \xi^{\beta} ),
\label{eq:current_B_gauge_trans}
\end{align}
which includes the diffeomorphism transformation, 
Yang-Mills gauge transformation, and the gauge transformation for the
anti-symmetric tensor field. 

The  C-bracket is obtained as the commutator of  $\Xi_i=\int d\sigma ~\xi_i^M\dd_M(\sigma)$ with $\xi_{12}{}^M=[\xi_1, \xi_2]_f^{M}$ 
in \bref{Cbra2}
	\bea
	[\Xi_1,\Xi_2]&=&i\Xi_{12}~~,~~\Xi_{12}=\int d\sigma ~\xi_{12}^M\dd_M(\sigma)~~.
	\eea
	The closure of the Jacobi identity of the $\Xi_1$, $\Xi_2$ and $V$ corresponds to
the equation given in  \bref{Cbra1}
	\bea
	[[\Xi_1, \Xi_2],V]+[[V,\Xi_1],\Xi_2]+[[\Xi_2,V],\Xi_1]&=&0~~,\nn\\
	\left[[\Xi_1, \Xi_2],V\right]&=&-\int d\sigma} \delta_{\xi_{12} V^M\dd_M~~,\nn\\
	\left[[V,\Xi_1],\Xi_2\right]+[[\Xi_2,V],\Xi_1]&=&\int d\sigma [\delta_{\xi_1},\delta_{\xi_2}]V^M\dd_M~~.
	\eea

In the last part of this section, we show that the (modified) field strength in heterotic
supergravities are obtained as a generalized torsion in the current algebras.
The commutator of covariant derivatives in the background gauge field $\dd_A=E_A{}^M\dd_M$ is given by
\bea
 [\dd_A(\sigma), \dd_B(\sigma')]&=&iT_{AB}{}^C\dd_C 
 \delta(\sigma-\sigma')+i\eta_{AB}  \partial_{\sigma} \delta(\sigma-\sigma')~~~. 
\eea
The torsion is given as
\bea
T_{AB}{}^M&=&E_{[A|}{}^N(\partial_N E_{|B]}{}^M)-E_A{}^N(\partial^M E_{BM})-
E_A{}^KE_B{}^Nf_{KN}{}^M, \nn\\
T_{ABC}&=&T_{AB}{}^{M}\eta_{MN}E_C{}^N=\frac{1}{3!}T_{[ABC]}, \nn\\
&=&\frac{1}{2}E_{[A|}{}^N(\partial_N
E_B{}^M)E_{|C]}{}^L\eta_{ML}-f_{ABC}, \nn \\
f_{ABC}&\equiv&E_A{}^ME_B{}^NE_C{}^Lf_{MNL}~.
\eea
In a physical gauge $\partial_M=(\partial_\mu,~\partial_\alpha=0,~\partial^\mu=0)$,
the commutators are given by
\bea
[\dd_m(\sigma), \dd_n(\sigma')]&=&i \left(\omega_{mn}{}^l\dd_l+
F_{mn}{}^c\dd_c+
\hat{H}
_{mnl}\dd^l \right) \delta(\sigma-\sigma'),~ \nn\\
\left[\dd_m(\sigma), \dd^n(\sigma')\right]&=&i \omega_{m}{}^n{}_l\dd^l\delta(\sigma-\sigma')
+i\delta_m^n\partial_\sigma\delta(\sigma-\sigma'),~\nn\\
\left[\dd^m(\sigma), \dd^n(\sigma')\right]&=&0,~\nn\\
\left[\dd_m(\sigma), \dd_b(\sigma')\right]&=&i \left(T_{mbc}\dd^c+F_{mnb}\dd^n \right)  \delta(\sigma-\sigma'),~\nn\\
\left[\dd_a(\sigma), \dd_b(\sigma')\right]&=&i\left(-f_{abc}\dd^c+T_{abl}\dd^l\right)\delta(\sigma-\sigma')
+i\kappa_{ab}\partial_\sigma\delta(\sigma-\sigma'),~\nn\\
\omega_{mn}{}^l&=&e_m{}^\mu e_n{}^\nu (\partial_{[\mu}e_{\nu]}{}^l), \nn\\
F_{mn}{}^c&=&e_m{}^\mu e_n{}^\nu F_{\mu\nu}{}^\gamma e_\gamma{}^c, \nn\\
F_{\mu\nu}{}^\gamma&=&\partial_{[\mu}A_{\nu]}{}^\gamma-A_{\mu}{}^\alpha
A_\nu{}^\beta f_{\alpha\beta\gamma}, \nn\\
\hat{H}_{mnl}
&=&e_m{}^\mu e_n{}^\nu e_l{}^\rho 
\hat{H}_{\mu \nu \rho}, \nn\\
\hat{H}_{\mu \nu \rho}
&=&\frac{1}{2}\left\{
\partial_{[\mu}B_{\nu\rho]}+
A_{[\mu}{}^\alpha
\partial_{\nu}
A_{\rho]
{}_\alpha}\right\}-
A_\mu{}^{\alpha} A_\nu{}^{\beta} A_{\rho}{}^\gamma f_{\alpha\beta\gamma}, ~~\nn\\
T_{mbc}&=&
\frac{1}{2}
e_a{}^\mu
(\partial_{\mu}e_{[b}{}^\beta)e_{c]\beta}-
e_a{}^\mu  A_\mu{}^\alpha 
e_b{}^\beta e_c{}^\gamma 
f_{\alpha\beta\gamma}
=T_{bcm},
\nn\\
f_{abc}
&=&
e_a{}^\alpha e_b{}^\beta e_c{}^\gamma
f_{\alpha\beta\gamma}
~.
\label{eq:torsion}
\eea 
We find the precise agreement of the expressions \eqref{eq:torsion} and
the quantities in \eqref{eq:heterotic_sugra_action}.

Finally, we comment on the last term in the heterotic supergravity
action \eqref{eq:heterotic_sugra_action}, namely, the curvature square
derivative corrections.
It is noteworthy that the gauge field $A_{\mu}^{~\alpha}$ and the spin
connection $\omega_{+\mu}^{~\Lambda}$ enter the supergravity Lagrangian
\eqref{eq:heterotic_sugra_action} in a completely symmetric way (cf. equations
\eqref{eq:Chern-Simons_terms}, \eqref{eq:sugra_gauge}).
By focusing on the parallel treatment of the gauge fields
$A_{\mu}^{~\alpha}$ and the spin connection $\omega_{\mu}^{~\Lambda}$,
the curvature square term is implemented as a part of the gauge kinetic
term in \cite{Bedoya:2014pma}.
This idea has been proposed in \cite{Polacek:2013nla} where the local
Lorentz symmetry is realized in the geometry with the extra coordinate.
The gauge space is extended to include the local Lorentz symmetry.
The extra coordinate is $Y^{\Lambda} = Y^{mn}$.
The local Lorentz transformation is realized as a gauge transformation
associated with the spin connection $\omega_{\mu}^{~\Lambda}$.

It is natural to extend our discussions
to that including the local Lorentz sector.
If the Riemann tensor is provided analogously to the gauge field strength 
in a way of the anomaly cancelation mechanism \cite{Green:1984sg},
then we recover the transformation \eqref{eq:current_B_gauge_trans} 
with redefinition of the gauge parameter \eqref{redef:xi}
in the leading order of the gauge coupling constant.
We may introduce further extra unphysical coordinates $Y^{\Lambda}$ and
enlarge the current algebra \eqref{affineLA} to include the Lorentz
sector.
We can write down the component transformation rules including the local
Lorentz transformations.
The torsion \eqref{eq:torsion} exhibits the modified field strength
$\hat{H}_{\mu \nu \rho}$ that contains Lorentz Chern-Simons term in
addition to the gauge Chern-Simons term.
Although this works well on the surface and is fine within the framework
of the leading order supergravity and DFT, 
it could cause potential problems in the higher orders in supergravity
and string theory.
This is because the local Lorentz group contains the translational symmetry
generated by $P_{\mu}$ and the structure constant $f^{\alpha} {}_{\beta
\gamma}$ for the Lorentz symmetry generically contains the spacetime indices.
It is apparent that when generators having spacetime indices such as 
the ones in the Poincare algebra and superalgebras live in the
internal gauge space, the condition $f^{\mu}{}_{NK}\partial_{\mu}\ast=0$ is not
appropriate since this implies all the fields are constant.

In the current algebra approach based on the non-degenerate Poincare
algebra, 
the generalized vielbein includes both the Lorentz connection and the
vielbein gauge field and the commutator of currents gives the Riemann tensor 
\cite{Polacek:2013nla}.
It is shown that the Jacobi identity requires the non-degenerate partner of the Lorentz generator. 
The concrete expression of currents were given where unphysical coordinates corresponding to the Lorentz and 
its non-degenerate partner do not introduce new physical degrees of
freedom and can be gauged away \cite{Hatsuda:2015cia}.

Since this discussion is convoluted and beyond the scope this paper, 
we leave the heterotic extension of this prescription for the future study.

\section{Conclusion and discussions} \label{sect:conclusion}

In this paper, we studied the structures of the $O(D,D+n)$
generalized metric and the non-Abelian gauge symmetries in the 
gauged DFT from the viewpoint of sigma models and current algebras.
We showed that the $O(D,D+n)$ generalized metric involving the gauge
sector in the gauged DFT is derived from a heterotic sigma models in arbitrary
backgrounds.
The sigma model has 480 extra unphysical chiral scalar fields
$Y^{\underline{\alpha}}$ in order to manifest the non-Abelian heterotic
gauge groups.
By taking these fields into account, we wrote down the heterotic sigma model in the curved background
including the non-Abelian gauge fields.
The extra coordinates $Y^{\underline{\alpha}}$ are needed to implement the gauge dynamics in
the heterotic theories.
They are parts of the doubled coordinate and are removed from the theory
by a constraint.
The heterotic sigma model derived in this paper is a non-Abelian generalization of
the one in \cite{Narain:1986am} where only the Cartan subsector of the gauge fields
are involved.

We then identified a stringy derivative operator in the heterotic sigma
models and utilized it to study the current algebras.
We calculated the gauge transformations of the generalized vielbein by
acting on the stringy derivative operators.
The transformation rules for the background fields are extracted from
the algebras.
We showed that they precisely reproduce the gauge
and the diffeomorphism transformations in the gauged DFT and hence
heterotic supergravities at order $\alpha'$.

We also discussed a possible extension including the local Lorentz
sector which is inevitable when the cancellation of anomaly and 
derivative corrections are taken
into account.
The Riemann curvature square term in heterotic supergravities, which is
a derivative correction, is treated as field strength of
the spin connection. As indicated in \cite{Bergshoeff:1988nn}, this is a completely parallel treatment to the
gauge sector.
Indeed in the DFT language, the local Lorentz symmetry is implemented as a part of the
doubled geometry in the same way as the gauge symmetries.
At least in $\mathcal{O}(\alpha')$, it has been shown that this works well \cite{Bedoya:2014pma}.
With this similarity, it is conceivable that the local Lorentz symmetry
and hence a derivative correction is incorporated with the sigma model
with extra gauge coordinates.

Our results revealed that the structures of the $\alpha'$- and the derivative
corrections and the non-Abelian gauge symmetries in the gauged DFT elucidated in \cite{Bedoya:2014pma} are
consistent with the heterotic sigma models.
In \cite{Bedoya:2014pma}, a relation between the $\alpha'$-corrections
in the generalized metric and the doubled $\alpha'$-geometry
\cite{Hohm:2013jaa, Hohm:2014xsa} is discussed.
It would be interesting to investigate this direction in the heterotic
sigma models.
The $O(D,D)$ T-duality symmetry of the stringy gravity theory  will provide a new description to the Planck scale physics, such as the early universe singularity, where winding modes play an essential role as well as momentum modes. 
Since the string sigma model with manifest T-duality describes not only massless modes but also massive modes,
it will be suitable for the string gas cosmology with manifest T-duality.
Furthermore the internal gauge symmetry is incorporated in its heterotic extension,
where the $O(D,D)$ covariance of the anomaly cancellation mechanism is interesting question.
 We will come back to this issue in the future work.

\subsection*{Acknowledgments}
We are grateful to Yuho Sakatani for informing us of helpful references
on DFT of heterotic string theories. 
This work is supported in part by Grant-in-Aid for Scientific Research (C), JSPS KAKENHI
Grant Numbers JP22K03603 (M.~H. and M.~Y.), JP20K03604 (M.~H.) and JP20K03952
(S.~S.).
The work of H~.M. is supported by Grant-in-Aid for JSPS Research Fellow, JSPS KAKENHI Grant Number JP22J14419 and JP22KJ2651.


\end{document}